\documentclass[12pt]{article}

\begin{document}

\newcommand{\siml}{\stackrel{<}{\sim}}
\newcommand{\simg}{\stackrel{>}{\sim}}
\newcommand{\lleq}{\stackrel{<}{=}}

\baselineskip=1.333\baselineskip


%
\begin{center}
{\large\bf
Effects of correlated variability on information entropies \\
in nonextensive systems
} 
\end{center}

\begin{center}
Hideo Hasegawa
\footnote{hideohasegawa@goo.jp}
\end{center}

\begin{center}
{\it Department of Physics, Tokyo Gakugei University  \\
Koganei, Tokyo 184-8501, Japan}
\end{center}
\begin{center}
({\today})
\end{center}
\thispagestyle{myheadings}

\begin{abstract}
We have calculated the Tsallis entropy and Fisher information matrix 
(entropy) of spatially-correlated nonextensive
systems, by using an analytic non-Gaussian distribution obtained
by the maximum entropy method.
Effects of the correlated variability on the Fisher information matrix
are shown to be different from those on the Tsallis entropy.
The Fisher information is increased (decreased) 
by a positive (negative) correlation,
whereas the Tsallis entropy is decreased with increasing 
an absolute magnitude of the correlation
independently of its sign.
This fact arises from the difference in their characteristics.
It implies from the Cram\'{e}r-Rao inequality
that the accuracy of unbiased estimate
of fluctuation is improved by the negative correlation.
A critical comparison is made between the present study and
previous ones employing the Gaussian approximation 
for the correlated variability due to multiplicative noise.

\end{abstract}

\vspace{0.5cm}
\noindent
PACS number(s): 05.70.-a,05.10.Gg,05.45.-a

\vspace{0.5cm}

\newpage

\section{Introduction}

It is well known that the Tsallis entropy and Fisher information entropy 
(matrix) are very important quantities expressing information measures
in nonextensive systems. The Tsallis entropy for $N$-unit nonextensive 
system is defined by
\cite{Tsallis88}-\cite{Tsallis04}
\begin{eqnarray}
S_q^{(N)} &=& \frac{(1-c_q^{(N)})}{(q-1)},
\label{eq:A1}
\end{eqnarray}
with
\begin{eqnarray}
c_q^{(N)} &=& \int [p^{(N)}(\{x_i \})]^q \:\Pi_i d x_i,
\label{eq:A2}
\end{eqnarray}
where $q$ is the entropic index ($0 < q < 3$), 
and $p^{(N)}(\{x_i \})$ denotes 
the probability distribution of $N$ variables $\{ x_i \}$.
In the limit of $q \rightarrow 1$, the Tsallis entropy
reduces to the Boltzman-Gibbs-Shannon entropy given by
\begin{eqnarray}
S_1^{(N)} &=& 
- \int p^{(N)}(\{x_i \}) \:\ln p^{(N)}(\{x_i \}) \: \Pi_i dx_i.
\label{eq:A3}
\end{eqnarray}
The Boltzman-Gibbs-Shannon entropy is extensive in the sense 
that for a system consisting $N$ independent but equivalent subsystems, 
the total entropy is a sum of constituent subsystems: 
$S_1^{(N)}=N S_1^{(1)}$. In contrast, the Tsallis entropy is nonextensive:
$S_q^{(N)} \neq N S_q^{(1)}$ for $q \neq 1.0$, and $\mid q-1 \mid$ 
expresses the degree of the nonextensivity of a given system.
The Tsallis entropy is a basis of the nonextensive statistical mechanics,
which has been successfully applied to a wide class of systems
including physics, chemistry, mathematics, biology, and others
\cite{Tsallis04}.

The Fisher information matrix provides us with an important measure 
on information \cite{{Frieden98}}. Its inverse expresses the lower 
bound of decoding errors for unbiased estimator in the Cram\'{e}r-Rao 
inequality. It denotes also the distance between the neighboring points 
in the Rieman space spanned by probability distributions in the information 
geometry. The Fisher information matrix expresses a local measure 
of positive amount of information whereas the Boltzman-Gibbs-Shannon-Tsallis entropy
represents a global measure of ignorance \cite{Frieden98}.
In recent years, many authors have investigated the Fisher information 
in nonextensive systems \cite{Plastino95}-\cite{Hasegawa08b}.
In a previous paper \cite{Hasegawa08b}, 
we have pointed out that two types of {\it generalized} and 
{\it extended} Fisher information matrices are necessary for nonextensive 
systems \cite{Hasegawa08b}.
The generalized Fisher information matrix $g_{ij}^{(N)}$
obtained from the generalized Kullback-Leibler divergence
in conformity with the Tsallis entropy, is expressed by
\begin{eqnarray}
g_{ij}^{(N)} &=& q E\left[
\left( \frac{\partial \ln p^{(N)}(\{x_i \})}{\partial \theta_i}  \right)
\left( \frac{\partial \ln p^{(N)}(\{x_i \})}{\partial \theta_j}  \right)
\right],
\label{eq:A4}
\end{eqnarray}
where $E[\cdot\cdot\cdot]$ denotes the average over 
$p^{(N)}(\{x_i \})$ [$=p^{(N)}(\{x_i \}; \{ \theta_k \})$]
characterized by a set of parameters $\{ \theta_k \}$.
On the contrary, the extended Fisher information matrix $\tilde{g}_{ij}^{(N)}$
derived from the Cram\'{e}r-Rao inequality in nonextensive systems, 
is expressed by \cite{Hasegawa08b}
\begin{eqnarray}
\tilde{g}_{ij}^{(N)} &=&  
E_q\left[
\left( \frac{\partial \ln P_q^{(N)}(\{x_i \})}{\partial \theta_i}  \right)
\left( \frac{\partial \ln P_q^{(N)}(\{x_i \})}{\partial \theta_j}  \right)
\right],
\label{eq:A5}
\end{eqnarray}
where $E_q[\cdot\cdot\cdot]$ expresses the average over the escort 
probability $P^{(N)}_q(\{x_i \} )$ given by
\begin{eqnarray}
P^{(N)}_q(\{x_i \} ) &=& \frac{[p^{(N)}(\{ x_i\})]^q}{c_q^{(N)}},
\end{eqnarray}
$c_q^{(N)}$ being given by Eq. (\ref{eq:A2}). In the limit of $q=1.0$, 
both the generalized and extended Fisher information matrices reduce to the 
conventional Fisher information matrix.

Studies on the information entropies have been made mainly for
independent (uncorrelated) systems. Effects of correlated noise and 
inputs on the Fisher information matrix and Shannon's mutual information 
have been extensively studied in neuronal ensembles (for a recent review, 
see Ref. \cite{Averbeck06}; related references therein).
It is a fundamental problem in neuroscience to determine whether 
correlations in neural activity are important for decoding,
and what is the impact of correlations on information transmission.
When neurons fire independently, the Fisher information increases
proportionally to the population size. In ensembles with the 
limited-range correlations, however, the Fisher information 
is shown to saturate as a function of population size
\cite{Abbott99}-\cite{Wu04}. 
In recent years the interplay between fluctuations and correlations
in nonextensive systems has been investigated 
\cite{Tsallis05}-\cite{Wilk07}.
It has been demonstrated that in some globally correlated
systems, the Tsallis entropy becomes extensive while
the Boltzman-Gibbs-Shannon entropy is nonextensive \cite{Tsallis05}.
Thus the correlation plays important roles in discussing the properties 
of information entropies in nonextensive systems.

It is the purpose of the present paper to study effects of the
spatially-correlated variability on the Tsallis entropy and Fisher information 
in nonextensive systems. 
In Sec. 2, we will discuss information entropies of correlated 
nonextensive systems, by using the probability distributions derived 
by the maximum entropy method (MEM). In Sec. 3, we discuss the 
marginal distribution to study the properties of probability 
distributions obtained by the MEM. Previous related studies are 
critically discussed also. The final Sec. 4 is devoted to our conclusion.
In appendix A, results of the MEM for uncorrelated,
nonextensive systems are briefly summarized 
\cite{Tsallis95,Borland99,Plastino00,Hasegawa08b,Hasegawa06}.

\section{Correlated nonextensive systems}

\subsection{The case of $N=2$}

We consider correlated $N$-unit nonextensive systems, 
for which the probability distribution is derived 
with the use of the MEM under the constraints given by
\begin{eqnarray}
1 &=& \int p^{(N)}(\{ x_i \})\: \Pi_i dx_i, 
\label{eq:C21}
\\
\mu &=& \frac{1}{N}\sum_i E_q\left[ x_i \right], 
\label{eq:C22}\\
\sigma^2 &=& \frac{1}{N} \sum_i
E_q\left[(x_i-\mu)^2 \right], 
\label{eq:C23} \\
s \:\sigma^2 &=& \frac{1}{N(N-1)}\sum_i \sum_{j (\neq i)}
E_q\left[(x_i-\mu)(x_j-\mu) \right],
\label{eq:C24}
\end{eqnarray}
$\mu$, $\sigma^2$ and $s$ expressing the mean, variance, and
degree of the correlated variability, respectively.
Cases with $N=2$ and arbitrary $N$ will be separately discussed 
in Secs. 2.1 and 2.2, respectively.

For a given correlated nonextensive system with $N=2$, the MEM with 
constraints given by Eqs. (\ref{eq:C21})-(\ref{eq:C24}) 
yields (details being explained in appendix B)
\begin{eqnarray}
p^{(2)}(x_1,x_2) &=& \frac{1}{Z_q^{(2)}}
\exp_q\left[- \left( \frac{1}{2}\right) 
\sum_{i=1}^{2} \sum_{j=1}^{2} A_{ij} 
(x_i-\mu)(x_j-\mu) \right],
\label{eq:C5}
\end{eqnarray}
with
\begin{eqnarray}
A_{ij} &=& a \:\delta_{ij}+ b \:(1-\delta_{ij}), 
\label{eq:C6}\\
a &=& \frac{1}{\nu_q^{(2)} \sigma^2(1-s^2)}, 
\\
b &=& - \frac{s}{\nu_q^{(2)}\sigma^2(1-s^2)}, 
\\
Z_q^{(2)}&=& \frac{2 \nu_q^{(2)} \sigma^2 r_q^{(2)}}{(q-1)}
B\left(\frac{1}{2},\frac{1}{q-1}-\frac{1}{2} \right)
B\left(\frac{1}{2},\frac{1}{q-1}-1 \right),
\hspace{0.5cm}\mbox{for $1< q < 3$}  
\label{eq:C9} \\
&=& 2 \pi \sigma^2 r_q^{(2)},
\hspace{7cm}\mbox{for $ q = 1$} 
\label{eq:C10} \\
&=& \frac{2 \nu_q^{(2)} \sigma^2 r_q^{(2)}}{(1-q)}
B\left(\frac{1}{2},\frac{1}{1-q}+1 \right)
B\left(\frac{1}{2},\frac{1}{1-q}+\frac{3}{2} \right),
\hspace{0.5cm}\mbox{for $0< q < 1$} 
\label{eq:C11}\\
r_q^{(2)} &=& \sqrt{1-s^2}, 
\label{eq:C8} \\
\nu_q^{(2)} &=& (2-q),
\label{eq:C12}
\end{eqnarray}
where 
$B(x,y)$ denotes the beta function
and $\exp_q(x)$ expresses the $q$-exponential function
defined by 
\begin{equation}
\exp_q(x) \equiv [1+(1-q)x]^{1/(1-q)}.
\label{eq:C13}
\end{equation}
The matrix ${\sf A}$ with elements $A_{ij}$ is expressed by the inverse 
of the covariant matrix ${\sf Q}$ given by
\begin{eqnarray}
{\sf A} &=& {\sf Q}^{-1},
\label{eq:D1} 
\end{eqnarray}
with
\begin{eqnarray}
Q_{ij} &=& \nu_q^{(2)} \sigma^2[ \delta_{ij}+s (1-\delta_{ij})]. 
\hspace{1cm}\mbox{for $i,j=1,2$}
\label{eq:D2} 
\end{eqnarray}
In the limit of $q=1.0$, the distribution $p^{(2)}(x_1,x_2)$ reduces to
\begin{eqnarray}
p^{(2)}(x_1,x_2) &=& \frac{1}{2 \pi \sigma^2 \sqrt{1-s^2}}
\exp \left[- \left(\frac{1}{2} \right) \sum_{i=1}^2 \sum_{j=1}^2
(x_i-\mu)({\sf Q}^{-1})_{ij}(x_j-\mu) \right],
\label{eq:D3} 
\end{eqnarray}
which is nothing but the Gaussian distribution for $N=2$.

We have calculated information entropies, by using the distribution 
given by Eq. (\ref{eq:C5}).

\vspace{0.5cm}
\noindent
{\bf Tsallis entropy}

We obtain
\begin{eqnarray}
S_q^{(2)} &=&[1+\log(2 \pi \sigma^2)]+\log(r_q^{(2)}),
\hspace{1cm}\mbox{for $q=1$} \\
&=& \frac{1-c_q^{(2)}}{q-1},
\hspace{4cm}\mbox{for $q \neq 1$}
\label{eq:D6}
\end{eqnarray}
with 
\begin{eqnarray}
c_q^{(2)} &=& \nu_q^{(2)} (Z_q^{(2)})^{1-q},
\label{eq:D7}
\end{eqnarray}
where $Z_q^{(2)}$ is given by Eq. (\ref{eq:C9})-(\ref{eq:C11}).
From $r_q^{(2)}$ given by Eq. (\ref{eq:C8}), we may obtain the $s$ 
dependence of $c_q^{(2)}$ as given by
\begin{eqnarray}
c_q^{(2)}(s) &=& c_q^{(2)}(0)(1-s^2)^{(1-q)/2},
\label{eq:D8}\\
&\simeq & c_q^{(2)}(0)\left[ 1+\frac{(q-1)}{2}s^2 \right],
\hspace{2cm}\mbox{for $\mid s \mid \ll 1$}
\label{eq:D9}
\end{eqnarray}
which yields
\begin{eqnarray}
S_q^{(2)}(s) &\simeq& 
S_q^{(2)}(0)-\frac{c_q^{(2)}(0)}{2}s^2.
\hspace{2cm}\mbox{for $\mid s \mid \ll 1$}
\label{eq:D10}
\end{eqnarray}

Figure \ref{figH}(a) shows $S_q^{(N)}/N$ as a function 
of the correlation $s$ for $N=2$
(values of $\mu=0.0$ and $\sigma^2=1.0$ are hereafter adopted in
model calculations shown in Figs. \ref{figH}-\ref{figE}).
We note that Tsallis entropy is decreased with increasing 
absolute value of $s$ independently of its sign.

\vspace{0.5cm}
\noindent
{\bf Fisher information}

By using Eqs. (\ref{eq:A4}) and (\ref{eq:A5}) for $\theta_i=\theta_j=\mu$,
we obtain the Fisher information matrices given by
\begin{eqnarray}
g_q^{(2)} &=& \frac{2}{\sigma^2 (1+s)},
\label{eq:D12} \\
\tilde{g}_q^{(2)} &=& \frac{2q}{(2q-1)(2-q)\sigma^2 (1+s)},
\label{eq:D13}
\end{eqnarray}
which show that $g_q^{(2)}$ is independent of $q$
and that the inverses of both matrices are proportional 
to $\sigma^2 (1+s)$.

Figure \ref{figJ} shows the $s$ dependence of the 
extended Fisher information for $N=2$, whose
inverse is increased (decreased) for a positive (negative) $s$, 
depending on a sign of $s$ in contrast to $S_q$. 

\subsection{The case of arbitrary $N$}

It is possible to extend our approach to the case of arbitrary $N$, 
for which the MEM with the constraints given by 
Eqs. (\ref{eq:C21})-(\ref{eq:C24}) lead to the 
distribution given by (details being given in appendix B)
\begin{eqnarray}
p^{(N)}(\{x_i\}) &=& \frac{1}{Z_q^{(N)}}
\exp_q\left[- \left( \frac{1}{2}\right)
\sum_{i=1}^{N} \sum_{j=1}^{N} A_{ij} 
(x_i-\mu)(x_j-\mu) \right],
\label{eq:F1}
\end{eqnarray}
with
\begin{eqnarray}
A_{ij} &=& a \;\delta_{ij}+ b \:(1-\delta_{ij}), 
\label{eq:F0} \\
a &=& \frac{[1+(N-2)s]}{\nu_q^{(N)}\sigma^2(1-s)[1+(N-1)s]}, \\
b &=& - \:\frac{s}{\nu_q^{(N)}\sigma^2(1-s)[1+(N-1)s]}, \\
%
%
Z_q^{(N)}
&=& \frac{(2 \nu_q^{(N)}  \sigma^2)^{N/2} \: r_q^{(N)}}
{(q-1)^{N/2}}\;\; \Pi_{i=1}^N 
\:B\left(\frac{1}{2}, \frac{1}{q-1}-\frac{i}{2} \right),
\hspace{1cm} \mbox{for $1 < q <3$}\\
&=& (2 \pi  \sigma^2)^{N/2} \: r_q^{(N)},
\hspace{6cm} \mbox{for $q=1$} \\
&=& \frac{(2 \nu_q^{(N)} \sigma^2)^{N/2} \: r_q^{(N)}}
{(1-q)^{N/2}}\;\; \Pi_{i=1}^N 
\:B\left(\frac{1}{2}, \frac{1}{1-q}+\frac{(i+1)}{2} \right),
\hspace{0.5cm} \mbox{for $0 < q <1$}  \\
r_q^{(N)} &=& \: \{(1-s)^{N-1}[1+(N-1)s] \}^{1/2},
\label{eq:F3} \\ 
\nu_q^{(N)} &=& \frac{[(N+2)-Nq]}{2}.
\label{eq:F4}
\end{eqnarray}
The matrix ${\sf A}$ is expressed by the inverse 
of the covariant matrix ${\sf Q}$ whose elements are given by
\begin{eqnarray}
Q_{ij} &=& \nu_q^{(N)} \sigma^2[ \delta_{ij}+s (1-\delta_{ij})]. 
\end{eqnarray}
In the limit of $q=1.0$, the distribution given by Eq. (\ref{eq:F1}) 
becomes the multivariate Gaussian distribution given by
\begin{eqnarray}
p(\{ x_i \}) &\propto&
\exp\left[-\left(\frac{1}{2}\right) 
\sum_{ij} (x_i-\mu)({\sf Q}^{-1})_{ij}(x_j-\mu) \right].
\end{eqnarray}

It is necessary to note that there is the condition for a 
physically conceivable 
$s$ value given by [see Eq. (\ref{eq:Z6}), details being discussed 
in appendix C]
\begin{equation}
s_L \leq s \leq s_U,
\label{eq:F8}
\end{equation}
where the lower and upper critical $s$ values are 
given by $s_L = -1/(N-1)$ 
and $s_U=1.0$, respectively. In the case of $N=2$ and $N=10$,  
for example, we obtain $s_L=-1.0$ and $s_L=-0.11$, respectively.

By using the probability distribution given by Eq. (\ref{eq:F1}),
we have calculated information entropies 
whose $s$ dependences are given as follows.

\vspace{0.5cm}
\noindent
{\bf Tsallis entropy}

We obtain
\begin{eqnarray}
S_q^{(N)} &=& \frac{N}{2}[1+\log(2 \pi \sigma^2)]+\log(r_q^{(N)}),
\hspace{1cm}\mbox{for $q=1$} \\
&=& \frac{1-c_q^{(N)}}{q-1},
\hspace{4cm}\mbox{for $q \neq 1$}
\label{eq:F5b}
\end{eqnarray}
with
\begin{eqnarray}
c_q^{(N)} &=& \nu_q^{(N)} (Z_q^{(N)})^{1-q}, 
\label{eq:F5}\\
&\simeq & c_q^{(N)}(0)\left[ 1+\frac{(q-1)N(N-1)}{4}s^2 \right],
\hspace{0.5cm}\mbox{for $\mid s \mid \ll 2/\sqrt{N(N-1)} $}
\label{eq:F5c}
\end{eqnarray}
where 
the $s$ dependence of $c_q^{(N)}$ arises from a factor of $r_q^{(N)}$ 
in Eq. (\ref{eq:F3}), and $c_q^{(N)}(0)$ expresses the $s=0.0$ value 
of $c_q^{(N)}$. Equation (\ref{eq:F5}) yields the $s$ dependent 
$S_q^{(N)}$ given by
\begin{eqnarray}
S_q^{(N)}(s)
&\simeq& S_q^{(N)}(0)-\left[\frac{ N(N-1) c_q^{(N)}(0)}{4} \right]  s^2, 
\hspace{0.5cm}\mbox{for $\mid s \mid \ll 2/\sqrt{N(N-1)} $} 
\label{eq:F6}
\end{eqnarray}
where $S_q^{(N)}(0)$  stands for the Tsallis entropy for $s=0.0$.
The region where Eqs. (\ref{eq:F5c}) and (\ref{eq:F6}) hold becomes 
narrower for larger $N$.

The $s$ dependence of $S_q^{(N)}/N$ for $N=10$ is shown in
Fig. \ref{figH}(b), where
$S_q^{(N)}/N$ has a peak 
at $s=0.0$ and it is decreased with increasing $\mid s \mid$.
Comparing Fig. \ref{figH}(b) with Fig. \ref{figH}(a), we
notice that $s$-dependence of $S_q^{(N)}/N$ 
for $N=10$ is more significant than that for $N=2$ [Eq. (\ref{eq:F6})].

Circles in Figs. \ref{figF}(a) and \ref{figF}(b) show $S_q^{(N)}/N$
with $s=0.0$ 
for $q < 1.0$ and $q > 1.0$, respectively, which are 
calculated with the use of the expressions given by Eqs. (\ref{eq:F5b}) 
and (\ref{eq:F5}). They are in good agreement with dashed curves showing
exact results which are given by Eq. (\ref{eq:B5}) 
and shown in Figs. \ref{figE}(a) and \ref{figE}(b)
in appendix A.
Squares show $S_q^{(N)}/N$ with $s=0.5$
calculated by using Eqs. (\ref{eq:F5b}) and (\ref{eq:F5}).
The Tsallis entropy is decreased by an introduced correlation.
Because of a computational difficulty \cite{Note2}, calculations using 
Eqs. (\ref{eq:F5b}) and (\ref{eq:F5}) cannot be performed for larger $N$ 
than those shown in Figs. \ref{figF}(a) and \ref{figF}(b).

\vspace{0.5cm}
\noindent
{\bf Fisher information}

The generalized and extended Fisher information matrices are given by
\begin{eqnarray}
g_q^{(N)} &=& \frac{N}{\sigma^2 [1+(N-1) s]}, 
\label{eq:F10} \\ 
\tilde{g}_q^{(N)} &=& \frac{Nq(q+1)}
{\sigma^2(3-q)(2q-1) [1+(N-1) s]}.
\label{eq:F11}
\end{eqnarray}
The results for $q=1.0$ given by Eqs. (\ref{eq:F10}) and (\ref{eq:F11}) 
are consistent with those derived with the use of the multivariate 
Gaussian distribution \cite{Abbott99}. 
By using the Fisher information matrices for $N=1$,
$g_q^{(1)}$ and $\tilde{g}_q^{(1)}$, given by 
Eqs. (\ref{eq:B9a}) and (\ref{eq:B10a}),
we obtain
\begin{eqnarray}%
\frac{g_q^{(1)}}{g_q^{(N)}} 
= \frac{\tilde{g}_q^{(1)}}{\tilde{g}_q^{(N)}} 
&=& \frac{1}{N}+\left( 1-\frac{1}{N} \right) s,
\label{eq:F7} \\
&=& s, 
\hspace{1cm}\mbox{for $N \rightarrow \infty$},
\label{eq:F14} \\
&=& \frac{1}{N}, 
\hspace{1cm}\mbox{for $s=0$}, \label{eq:F13} \\
&=& 1, 
\hspace{1cm}\mbox{for $s=s_U$} \\
&=& 0, 
\hspace{1cm}\mbox{for $s=s_L$}
\label{eq:F9}
\end{eqnarray}
which holds independently of $q$.
Inverses of the Fisher information matrices approach 
the value of $s$ for $N \rightarrow \infty$, and
are proportional to $1/N$ for $s=0.0$.  
In particular, they vanish at $s=s_L$.  
These features are clearly seen in
Fig. \ref{figD} where the inverses of Fisher information matrices,
$g_q^{(1)}/g_q^{(N)}$ and $\tilde{g}_q^{(1)}/\tilde{g}_q^{(N)}$,
are plotted as a function of $N$ for various $s$ values. 

\section{Discussion}

\subsection{Marginal distributions}

In the present study, we have obtained the probability 
distributions, applying the MEM to spatially-correlated nonextensive systems.
We will examine our probability distributions in more detail. 
The $x_1$ dependences of $p^{(2)}(x_1, x_2)$ for $N=2$ 
given by Eq. (\ref{eq:C5}) with $s=0.0$ and $s=0.5$ are plotted in 
Figs. \ref{figC}(a) and \ref{figC}(b), respectively, where $x_2$ is 
treated as a parameter. When $s=0.0$, the distribution is symmetric 
with respect of $x_1$ for all $x_2$ values. 
When the correlated variability of $s=0.5$ is introduced,
peak positions of the distribution appear at finite $x_1$
for $x_2=0.5$ and 1.0.

In the limit of $s=0.0$ ({\it i.e.} no correlated variability), 
$p^{(2)}(x_1, x_2)$ given by Eq. (\ref{eq:C5}) becomes
\begin{eqnarray}
p^{(2)}(x_1,x_2) &\propto& 
\left[ 1-\frac{(1-q)(x_1^2+x_2^2)}{2 \nu_q^{(2)}\sigma^2}
\right]^{1/(1-q)},
\label{eq:G8}
\end{eqnarray}
which does not agree with the exact result (except for $q=1.0$),
as given by
\begin{eqnarray}
p^{(1)}(x_1)p^{(1)}(x_2)
&\propto& \left[ 1-\frac{(1-q)(x_1^2+x_2^2)}{ 2 \nu_q^{(1)}\sigma^2}
+ \frac{(1-q)^2}{4 (\nu_q^{(1)})^2\sigma^4} x_1^2 x_2^2 \right]^{1/(1-q)},
\label{eq:G} \\
&\neq & p^{(2)}(x_1,x_2),
\end{eqnarray}
because of the properties of
the $q$-exponential function defined by Eq. (\ref{eq:C13}):
$\exp_q(x+y) \neq \exp_q(x)\exp_q(y)$.
By using the $q$-product $\otimes_q$ defined by
\cite{Borges04}
\begin{eqnarray}
x \otimes_q y \equiv [x^{1-q}+y^{1-q}-1]^{1/(1-q)}, 
\end{eqnarray}
we may obtain the expression given by
\begin{eqnarray}
p^{(1)}(x_1) \otimes_q p^{(1)}(x_2) &\propto& 
\left[ 1-\frac{(1-q)(x_1^2+x_2^2)}{2 \nu_q^{(1)}\sigma^2}
\right]^{1/(1-q)},
\label{eq:G16}
\end{eqnarray}
which coincides with $p^{(2)}(x_1,x_2)$ given by Eq. (\ref{eq:G8}) 
besides a difference between $\nu_q^{(1)}$ and $\nu_q^{(2)}$.

In order to study the properties of the probability distribution
of $p^{(2)}(x_1,x_2)$ in more details, 
we have calculated its marginal probability 
(with $s=0.0$) given by 
\begin{eqnarray}
p_m^{(2)}(x_1) &=& \int p^{(2)}(x_1,x_2)\: dx_2
\propto 
\left[ 1- \frac{(1-q)x_1^2}
{2 \nu_q^{(2)}\sigma^2} \right]^{1/(1-q)+1/2}.
\label{eq:G4}
\end{eqnarray}
Dashed curves in Figs. \ref{figB}(a) and \ref{figB}(b) show 
$p_m^{(2)}(x_1)$ in linear and logarithmic scales, respectively.
The marginal distributions are in good agreement with solid curves
showing $p^{(1)}(x_1)$ [Eq. (\ref{eq:X4})],
\begin{eqnarray}
p^{(1)}(x_1) &\propto&
\left[ 1- \frac{(1-q)x_1^2}{2 \nu_q^{(1)}\sigma^2} \right]^{1/(1-q)}.
\label{eq:G7}
\end{eqnarray}
In the case of $N=3$, the distribution given by Eq. (\ref{eq:F1}) yields 
its marginal distribution (with $s=0.0$) given by
\begin{eqnarray}
p_m^{(3)}(x_1) &=& \int \int p^{(3)}(x_1,x_2,x_3)\: dx_2\:dx_3
\propto 
\left[ 1- \frac{(1-q)x_1^2}
{2 \nu_q^{(3)}\sigma^2} \right]^{1/(1-q)+1}.
\label{eq:G6}
\end{eqnarray}
Chain curves in Figs. \ref{figB}(a) and \ref{figB}(b) represent
$p_m^{(3)}(x_1)$, which is again in good agreement 
with solid curves showing $p^{(1)}(x_1)$. These results justify, 
to some extent, the probability distribution adopted 
in our calculation.

The marginal distribution for an arbitrary $N$ (with $s=0.0$) is given by
\begin{eqnarray}
p_m^{(N)}(x_1) &=& 
\int \int p^{(N)}(x_1,\cdot\cdot, x_N)\: dx_2\:\cdot\:\cdot\:dx_N \\
&\propto& 
\left[ 1- \frac{(1-q)x_1^2}
{2 \nu_q^{(N)}\sigma^2} \right]^{1/(1-q)+(N-1)/2}, \\
&\propto&
\left[ 1- \frac{(1-q_N) x_1^2}
{2 \nu_N \sigma^2} \right]^{1/(1-q_N)}, 
\label{eq:G11}
\end{eqnarray}
with
\begin{eqnarray}
q_N &=& \frac{(N-1)-(N-3)q}{(N+1)-(N-1)q}, 
\label{eq:G12} \\
\nu_N & = & \frac{(N+2)-N q}{(N+1)-(N-1)q}. 
\label{eq:G13} 
\end{eqnarray}
Equations (\ref{eq:G11})-(\ref{eq:G13}) show that in the limit of 
$N \rightarrow \infty$, we obtain $q_N = 1.0$, $\nu_N=1.0$,
and $p_m^{(N)}(x_1)$ reduces to the Gaussian distribution.

\subsection{Comparison with related studies}

One of typical microscopic nonextensive systems is the Langevin model 
subjected to multiplicative noise, as given by 
\cite{Hasegawa07}-\cite{Anten02}
\begin{eqnarray}
\frac{dx_i}{dt}\!\!&=&\!\! - \lambda x_i + \beta \xi_i(t)
+ \alpha \: x_i\: \eta_i(t)+H(I).
\hspace{1cm}\mbox{($i=1$-$N$)}
\label{eq:G10} 
\end{eqnarray}
Here $\lambda$ expresses the relaxation rate,
$H(I)$ denotes a function of an external input $I$, 
and $\alpha$ and $\beta$ stand for magnitudes of
multiplicative and additive noise, respectively,
with zero-mean white noise given by $\eta_i(t)$ and $\xi_i(t)$
with the correlated variability,
\begin{eqnarray}
\langle \eta_i(t)\:\eta_j(t') \rangle 
&=& \alpha^2 [\delta_{ij}+ c_M (1-\delta_{ij})] \delta(t-t'),\\
\langle \xi_i(t)\:\xi_j(t') \rangle 
&=& \beta^2 [\delta_{ij} + c_A(1-\delta_{ij})]\delta(t-t'),\\
\langle \eta_i(t)\:\xi_j(t') \rangle &=& 0,
\label{eq:H1}
\end{eqnarray}
where $c_A$ and $c_M$ express the degrees of correlated variabilities
of additive and multiplicative noise, respectively.
The Fokker-Planck equation (FPE) 
for the probability distribution $p(\{ x_k\},t)$ ($=p$)
is given by
\begin{eqnarray}
\frac{\partial}{\partial t}\: p &=&
-\sum_{i} \frac{\partial}{\partial x_{i}}[ (-\lambda x_i + H)
\:p ] 
\nonumber \\
&+&\frac{\beta^2}{2}\sum_{i}\sum_{j}
[ \delta_{ij} + c_A (1-\delta_{ij}) ]
\frac{\partial^2}{\partial x_{i} \partial x_{j}} \:p
\nonumber \\
&+&\frac{\alpha^2}{2}\sum_{i} \sum_j
[ \delta_{ij}+ c_M (1-\delta_{ij}) ] 
\frac{\partial}{\partial x_{i}} x_i
\frac{\partial}{\partial x_j}
(x_j \:p),
\label{eq:H0}
\end{eqnarray}
in the Stratonovich representation. 

For additive noise only ($\alpha=0$), the stationary distribution 
is given by
\begin{eqnarray}
p(\{ x_i \}) &\propto&
\exp[-\frac{1}{2} \sum_{ij} (x_i-\mu_i)({\sf Q}^{-1})_{ij}(x_j-\mu_j)],
\end{eqnarray}
where $\mu_i$ denotes the average of $x_i$, and ${\sf Q}$ expresses
the covariance matrix given by
\begin{eqnarray}
Q_{ij} &=& \beta^2 [\delta_{ij}+ c_A (1-\delta_{ij})].
\end{eqnarray}

When multiplicative noise exists $(\alpha \neq 0.0)$, the calculation 
of even stationary distribution becomes difficult, and it is generally 
not given by the Gaussian. Indeed, the stationary distribution 
for non-correlated multiplicative noise with
$\alpha \neq 0.0$, $\beta \neq 0.0$ and $c_A=c_M=0.0$ 
is given by 
\cite{Hasegawa08b}\cite{Hasegawa07}-\cite{Anten02}
\begin{eqnarray}
p(\{ x_i \}) &\propto& \Pi_i
\left[1-(1-q)\left( \frac{x_i^2}{2 \phi^2}\right) \right]^{1/(1-q)} 
\:e^{Y(x_i)},
\label{eq:H3}
\end{eqnarray}
with
\begin{eqnarray}
q &=& 1 + \frac{2 \alpha^2}{2 \lambda + \alpha^2}, \\
\phi^2 &=& \frac{\beta^2}{2 \lambda+ \alpha^2}, \\
Y(x_i) &=& \left( \frac{2\:H}{\alpha \beta} \right) 
\tan^{-1} \left( \frac{\alpha x_i}{\beta} \right).
\end{eqnarray}
The probability distribution given by Eq. (\ref{eq:H3})
for $H=0$ ($c_A=c_M=0$) agrees
with that derived by the MEM  
for $\phi^2= \nu_q^{(1)} \sigma^2$ [Eq. (\ref{eq:X4})]. 
For $\alpha \neq 0.0$, $\beta=0.0$ and $H=\mu > 0$ ($c_A=c_M=0$),
Eq. (\ref{eq:H3}) becomes \cite{Hasegawa08b}
\begin{eqnarray}
p(x) & \propto & \mid x \mid^{-2/(q-1)}
e^{-2 \mu/\alpha^2 x} \; \Theta(x), 
\label{eq:H6}  
\end{eqnarray}
yielding the Fisher information given by
\begin{eqnarray}
g_q^{(N)} &=& \frac{N q^4}{\sigma_q^2}
= \frac{2 N \lambda q^4}{\alpha^2 \mu^2},
\label{eq:H7} 
\end{eqnarray}
where $\sigma_q^2=\alpha^2 \mu^2/2 \lambda $ and
$\Theta(x)$ is the Heaviside function.

The probability distribution for correlated multiplicative noise
($\alpha \neq 0.0$, $c_M \neq 0.0$) is also the non-Gaussian, 
which is easily confirmed by direct simulations of the Langevin 
model with $N=2$ \cite{Hasegawa08c}.
In some previous studies \cite{Abbott99}-\cite{Wu04}, the stationary 
distribution of the Langevin model subjected to correlated multiplicative
noise with $c_M \neq 0.0$, $\beta=0.0$ and $H=\mu$ 
is assumed to be expressed by the Gaussian distribution
with the covariance matrix given by
\begin{eqnarray}
Q_{ij} &=& 
\sigma_M^2 \mu_i \:\mu_j[\delta_{ij}+ c_M (1-\delta_{ij})].
\label{eq:H2}
\end{eqnarray}
This is equivalent to assume that 
\begin{eqnarray}
\frac{\partial}{\partial x_i} \left[ x_i \frac{\partial}
{\partial x_j} (x_j \:p) \right]
&\simeq& \frac{\partial}{\partial x_i} 
\left[ \langle x_i \rangle  \frac{\partial}
{\partial x_j} (\langle x_j \rangle \:p) \right], \nonumber \\
&=& \mu_i \mu_j \frac{\partial^2 p}
{\partial x_i \partial x_j},
\label{eq:H3b}
\end{eqnarray}
in the FPE given  by Eq. (\ref{eq:H0}) with 
$\mu_i=\langle x_i \rangle $ and $\sigma_M^2=\alpha^2$. 
By using such an approximation,
Abbott and Dayan (AD) \cite{Abbott99} calculated the Fisher information
matrix of a neuronal ensemble with the correlated variability, 
which is given by
\begin{eqnarray}
g^{(N)}_{AD} &=& \frac{N K}{\sigma_M^2 [1+(N-1)c_M]}
+ 2 N K, 
\nonumber \\
&=& \frac{N}{\sigma_M^2 \mu^2 [1+(N-1)c_M]}+ \frac{2N}{\mu^2}, 
\label{eq:H4} 
\end{eqnarray}
with a spurious second term ($2 N K$),
where $K=N^{-1}\sum_i\:[d \ln H(\mu_i)/d \mu_i]^2=1/\mu^2$ 
[Eq. (4.7) of Ref. \cite{Abbott99} in our notation]. 
Equation (\ref{eq:H4}) 
is not in agreement with either 
Eq. (\ref{eq:F10}) or (\ref{eq:F11}) derived by the MEM.
Furthermore, the result of AD in the limit of $c_M=0$,  
$g^{(N)}_{AD}=(N/\sigma_M^2 \mu^2 +2N /\mu^2)$, 
does not agree with the exact result given by Eq. (\ref{eq:H7})
for the Langevin model.
This fact casts some doubt on the results of
Refs. \cite{Abbott99}-\cite{Wu04} based on
the Gaussian approximation given by Eq. (\ref{eq:H2})
or $(\ref{eq:H3b})$, which has no physical or mathematical justification.
The Fisher information matrix depends on 
a detailed structure of the probability distribution because
it is expressed by the derivative of the distribution  
with respect to $x$, as given by
\begin{eqnarray}
g_{q}^{(N)} &=& - q \:E\left[
\frac{\partial^2 \ln p^{(N)}(x)}{\partial x^2}
\right]. 
\label{eq:H5} 
\end{eqnarray}
We must take into account the non-Gaussian structure of the probability
distribution in discussing the Fisher information of nonextensive systems.

\section{Conclusion}

We have discussed effects of the spatially-correlated variability
on the Tsallis entropy and Fisher information 
matrix in nonextensive systems, 
by using the probability distribution derived by the MEM.
Although the obtained analytical distribution 
in the limit of $s=0.0$ does not hold
the relation given by 
$ p^{(N)}(\{ x_k\})= \Pi_{i=1}^N \; p^{(1)}(x_i)$ 
for $q \neq 1.0$,
it numerically yields good results (Fig. \ref{figF}), 
reducing to the multivariate Gaussian distribution
in the limit of $q=1.0$.
Our calculations have shown that

\noindent
(i) the Tsallis entropy is decreased by both positive and
negative correlations, and

\noindent
(ii) the inverses of the Fisher information matrices are increased (decrease)
by a positive (negative) correlation.

\noindent
The difference between the $s$ dependences of $S_q$
and $g_q$ arises from the difference in their characteristics:
$S_q$ provides us with a global measure of ignorance
while $g_q$ a local measure of positive amount of information
\cite{Frieden98}.
The item (ii) implies that the accuracy of
unbiased estimate of fluctuation is improved
by the negative correlation.
If there is known, and strong,
negative correlation between successive pairs of data, estimating the
unknown parameter as their simple average must reduce the error, as
negatively correlated errors tend to cancel in taking the difference.

In connection with the discussion presented in Sec. 3,
it is interesting to make a detailed study on the properties 
of information entropies in the Langevin model
subjected to correlated as well as uncorrelated multiplicative noise.
Such a calculation is in progress and will be reported elsewhere 
\cite{Hasegawa08c}.
 
\section*{Acknowledgments}
This work is partly supported by
a Grant-in-Aid for Scientific Research from the Japanese 
Ministry of Education, Culture, Sports, Science and Technology.  

\vspace{1.0cm}

\appendix

\noindent
{\large\bf Appendix A: MEM for non-correlated nonextensive systems}
\renewcommand{\theequation}{A\arabic{equation}}
\setcounter{equation}{0}

We summarize results of the MEM for nonextensive systems
\cite{Tsallis95,Borland99,Plastino00,Hasegawa08b,Hasegawa06}.
In order to apply the MEM to $N$-unit nonextensive systems,
we impose the three constraints given by 
\begin{eqnarray}
1 &=& \int p^{(N)}(\{ x_i \})\: \Pi_i dx_i, 
\label{eq:C1}
\\
\mu &=& \frac{1}{N}\sum_i E_q\left[ x_i \right], 
\label{eq:C2}\\
\sigma^2 &=& \frac{1}{N} \sum_i
E_q\left[(x_i-\mu)^2 \right],
\label{eq:C3}
\end{eqnarray}
where the $q$-dependent $\mu$ and $\sigma^2$ correspond to 
$\mu_q$ and $\sigma_q^2$, respectively, in \cite{Hasegawa08b}.
For a given nonextensive system, the variational condition 
for the Tsallis entropy given by Eq. (\ref{eq:A1}) with the constraints 
(\ref{eq:C1})-(\ref{eq:C3}) yields the $q$-Gaussian distribution given by
\cite{Tsallis95,Borland99,Plastino00,Hasegawa08b,Hasegawa06}

\begin{eqnarray}
p^{(N)}(\{x_k \}) &=& \Pi_{i=1}^N \: p^{(1)}(x_i),
\label{eq:B1}
\end{eqnarray}
with
\begin{eqnarray}
p^{(1)}(x) &=& \frac{1}{Z_q^{(1)}} 
\exp_q\left[- \left( \frac{(x-\mu)^2}
{2\nu_q^{(1)} \sigma^2}\right) \right],
\label{eq:X4}\\
Z_q^{(1)} &=& \int 
\exp_q\left(-\frac{(x-\mu)^2}{2 \nu_q^{(1)} \sigma^2} \right) \:dx, \\
&=& \left(\frac{2 \nu_q^{(1)}  \sigma^2}{q-1} \right)^{1/2}
B\left(\frac{1}{2}, \frac{1}{q-1}-\frac{1}{2} \right),
\hspace{1cm}\mbox{for $1< q < 3$}
\label{eq:X5} 
\\
&=& \sqrt{2 \pi} \sigma,
\label{eq:X6}
\hspace{5.5cm}\mbox{for $q=1.0$} 
\\
&=& \left(\frac{2 \nu_q^{(1)} \sigma^2}{1-q} \right)^{1/2}
B\left(\frac{1}{2}, \frac{1}{1-q}+1 \right),
\hspace{1cm}\mbox{for $0< q < 1$}
\label{eq:X7} \\
%
%
\nu_q^{(1)} &=& \frac{3-q}{2},
\label{eq:X9}
\end{eqnarray}
where 
$B(x,y)$ denotes the beta function.

\vspace{0.5cm}
\noindent
{\bf Tsallis entropy}

Substituting the probability distribution given by Eq. (\ref{eq:B1})
to Eqs. (\ref{eq:A1}) and (\ref{eq:A2}), we obtain the Tsallis entropy
expressed by
\begin{eqnarray}
S_q^{(N)} &=& \frac{N}{2}[1+\log(2 \pi \sigma^2)], 
\hspace{1cm}\mbox{for $q=1$} \\
&=& \frac{[1-(c_q^{(1)})^N]}{(q-1)},
\hspace{2cm}\mbox{for $q \neq 1$}
\label{eq:B3}
\end{eqnarray}
where $c_q^{(1)}=\nu_q^{(1)}(Z_q^{(1)})^{1-q}$.
We may express $S_q^{(N)}$ in terms of $S_q^{(1)}$ by
\cite{Hasegawa08b}
\begin{eqnarray}
S_q^{(N)} &=& \sum_{k=1}^N C_k^N 
(-1)^{k-1} (q-1)^{k-1} (S_q^{(1)})^k,
\label{eq:B5}  
\\
&=& N S_q^{(1)}-\frac{N(N-1)(q-1)}{2}(S_q^{(1)})^2+ \cdot\cdot\cdot, 
\label{eq:B6}
\end{eqnarray}
where $C_k^N=N!/(N-k)!\:k!$. Equation (\ref{eq:B6}) clearly shows that
the Tsallis entropy is generally nonextensive except for $q=1.0$ 
for which $S_q^{(N)}=N S_q^{(1)}$.
Figures \ref{figE}(a) and \ref{figE}(b) show the $N$ dependence of 
the Tsallis entropy per element, $S_q^{(N)}/N$, of uncorrelated 
systems ($s=0.0$), which are calculated with the use of 
Eq. (\ref{eq:B5}). With increasing $N$, $S_q^{(N)}/N$ is 
decreased for $q > 1.0$ whereas it is significantly increased for $q < 1.0$.

\vspace{0.5cm}
\noindent
{\bf The Fisher information}

With the use of Eqs. (\ref{eq:A4}) and (\ref{eq:A5})
for $\theta_i=\theta_j=\mu$,
the generalized and extended Fisher information matrices
are given by \cite{Hasegawa08b}
\begin{eqnarray}
g_q^{(N)}&=&N g_q^{(1)}, 
\label{eq:B9}\\
\tilde{g}_q^{(N)}&=&N \tilde{g}_q^{(1)}.
\label{eq:B10}
\end{eqnarray}
with
\begin{eqnarray}
g_q^{(1)}&=& \frac{1}{\sigma^2}, 
\label{eq:B9a}\\
\tilde{g}_q^{(1)}&=& \frac{q(q+1)}{(3-q)(2q-1)\sigma^2},
\label{eq:B10a}
\end{eqnarray}
which show that Fisher information matrices are extensive.

\vspace{0.5cm}

\noindent
{\large\bf Appendix B: MEM for correlated nonextensive systes}
\renewcommand{\theequation}{B\arabic{equation}}
\setcounter{equation}{0}

In the case of $N=2$, the probability distribution $p^{(2)}(x_1,x_2)$ 
given by Eqs. (\ref{eq:C5}) and (\ref{eq:C6}) is rewritten as
\begin{eqnarray}
p^{(2)}(x_1, x_2) &\propto& [1-(1-q) \Phi^{(2)}(x_1,x_2)]^{1/(1-q)},
\label{eq:Y1}
\end{eqnarray}
with
\begin{eqnarray}
\Phi^{(2)}(x_1,x_2) &=& \frac{1}{2}
\left[a\:(x_1^2+x_2^2)+2b \:x_1 x_2 \right],\\
&=& \lambda_1 y_1^2+\lambda_2 y_2^2, 
\label{eq:Y2}
\end{eqnarray}
where $\lambda_i$ and $y_i$ ($i=1,2$) are eigenvalues and eigenvectors 
of $\Phi(x_1,x_2)$ given by
\begin{eqnarray}
\lambda_1 &=& \frac{1}{2}(a+b), \\
\lambda_2 &=& \frac{1}{2}(a-b), \\
y_1 &=& \frac{1}{\sqrt{2}}(x_1+x_2), \\
y_2 &=& \frac{1}{\sqrt{2}}(x_1-x_2).
\end{eqnarray}
The averages of $E_q[y_1^2]$ and $E_q[(y_1^2]$ are
given by
\begin{eqnarray}
E_q[y_1^2] &=& \frac{1}{\nu_q^{(2)}\lambda_1}, \\
E_q[y_2^2] &=& \frac{1}{\nu_q^{(2)}\lambda_2},
\end{eqnarray}
from which we obtain $\sigma^2$ and $s\:\sigma^2$ as
\begin{eqnarray}
\sigma^2 &=& \frac{1}{2} E_q[x_1^2+x_2^2] 
=\frac{1}{2}E_q[y_1^2+y_2^2], \nonumber \\
&=& \frac{a}{\nu_q^{(2)}(a^2-b^2)}, 
\label{eq:Y3}\\
s\: \sigma^2 &=& E_q[x_1 x_2]
= \frac{1}{2}E_q[y_1^2-y_2^2], \nonumber \\
&=& - \frac{b}{\nu_q^{(2)}(a^2-b^2)}.
\label{eq:Y4}
\end{eqnarray}
By using Eqs. (\ref{eq:Y3}) and (\ref{eq:Y4}), $a$ and $b$ are expressed 
in terms of $\sigma^2$ and $s$ as
\begin{eqnarray}
a &=& \frac{1}{\nu_q^{(2)}\sigma^2 (1-s^2)}, \\
b &=& - \frac{s}{\nu_q^{(2)}\sigma^2 (1-s^2)},
\end{eqnarray}
which yield the matrix of ${\sf A}$ given by an inverse
of the covariance matrix of ${\sf Q}$ [Eq. (\ref{eq:D2})].

A calculation for the case of an arbitrary $N$ may be similarly 
performed as follows. The distribution given by Eqs. (\ref{eq:F1}) 
and (\ref{eq:F0}) is rewritten as
\begin{eqnarray}
p^{(N)}(\{ x_i \}) &\propto& [1-(1-q) \Phi^{(N)}(\{x_i \})]^{1/(1-q)},
\label{eq:W1}
\end{eqnarray}
with 
\begin{eqnarray}
\Phi^{(N)}(\{ x_i \}) &=& \frac{1}{2}
\left[a \sum_i x_i^2+ 2 b \sum_{i < j} \:x_i x_j \right],\\
&=& \sum_i \lambda_i y_i^2, 
\label{eq:W2}
\end{eqnarray}
where $\lambda_i$ and $y_i$ are eigenvalues and eigenvectors, 
respectively. With the use of eigenvalues given by
\begin{eqnarray}
\lambda_i &=& \frac{1}{2}[a+(N-1)b], 
\hspace{1cm}\mbox{for $i=1$} \\
&=& \frac{1}{2}(a-b), 
\hspace{1cm}\mbox{for $1 < i \leq N$}
\end{eqnarray}
Eqs. (\ref{eq:C23}) and (\ref{eq:C24}) lead to
\begin{eqnarray}
\sigma^2 &=& \frac{1}{N} \sum_i E_q[y_i^2], 
\nonumber \\
&=& \frac{1}{\nu_q^{(N)} N} \sum_i \left( \frac{1}{\lambda_i} \right),
\nonumber \\ 
&=& \frac{[a+b(N-2)]}{\nu_q^{(N)} (a-b)[a+(N-1)b]}, 
\label{eq:W3}\\
s\:\sigma^2 &=& \frac{1}{N(N-1)} \sum_{i<j}E_q[y_i^2-y_j^2], 
\nonumber\\
&=& \left( \frac{1}{\nu_q^{(N)} N(N-1)} \right)
\sum_{i < j}\left( \frac{1}{\lambda_i}-\frac{1}{\lambda_j} \right),
\nonumber\\
&=& -\: \frac{b}{\nu_q^{(N)} (a-b)[a+(N-1)b]}.
\label{eq:W4} 
\end{eqnarray}
From Eqs. (\ref{eq:W3}) and (\ref{eq:W4}), $a$ and $b$ are expressed 
in terms of $\sigma^2$ and $s$, as given by
\begin{eqnarray}
a &=& \frac{[1+(N-2)s]}{\nu_q^{(N)}\sigma^2(1-s)[1+(N-1)s]}, \\
b &=& - \:\frac{s}{\nu_q^{(N)}\sigma^2(1-s)[1+(N-1)s]}.
\end{eqnarray}

\vspace{1cm}
\noindent
{\large\bf Appendix C: The condition for a conceivable $s$ value}
\renewcommand{\theequation}{C\arabic{equation}}
\setcounter{equation}{0}

In order to discuss the condition for a physically conceivable $s$ value,
we consider the global variable $X(t)$ defined by
\begin{equation}
X(t) = \frac{1}{N} \sum_i x_i(t).
\label{eq:Z1}
\end{equation}
The first and second $q$ moments of $X(t)$ are given by
\begin{eqnarray}
E_q[X(t)] &=& \frac{1}{N} \sum_i E_q[x_i(t)] = \mu(t), \\
\label{eq:Z2}
E_q[\{ \delta X(t) \}^2] 
&=& \frac{1}{N^2} \sum_i \sum_j E_q[\delta x_i(t)\delta x_j(t)],\\
&=& \frac{1}{N^2} \sum_i E_q[\{ \delta x_i(t) \}^2]
+ \frac{1}{N^2} \sum_i \sum_{j (\neq i)} 
E_q[\delta x_i(t)\delta x_j(t)], \\
&=& \frac{\sigma(t)^2}{N}[1+(N-1)s(t)],
\label{eq:Z3}
\end{eqnarray}
where $\delta x_i=x_i(t)-\mu(t)$ and $\delta X(t)= X(t)-\mu(t)$.
Since global fluctuation in $X$ 
is smaller than the average of local fluctuation
in $\{ x_i \}$, we obtain
\begin{eqnarray}
0 \leq E_q[\{ \delta X(t) \}^2] \leq 
\frac{1}{N}\sum_i E[\{ \delta x_i(t)\}^2]=\sigma(t)^2.
\label{eq:Z4}
\end{eqnarray}
Equations (\ref{eq:Z3}) and (\ref{eq:Z4}) yield
\begin{equation}
0 \leq \frac{[1+(N-1)s(t)]}{N} \leq 1.0,
\label{eq:Z5}
\end{equation}
which leads to 
\begin{equation}
s_L \leq s(t) \leq s_U,
\label{eq:Z6}
\end{equation}
with $s_L=-1/(N-1)$ and $s_U=1.0$.


\newpage

\begin{figure}
\begin{center}
\end{center}
\caption{
The $s$ dependence of the Tsallis entropy per element,
$S_q^{(N)}/N$, with (a) $N=2$ and (b) $N=10$
for various $q$ values with $\mu=0.0$ and $\sigma^2=1.0$:
the $s$ value is allowed to be $-1.0 <s < 1.0$ for $N=2$,
and $-0.11 <s < 1.0$ for $N=10$ [Eq. (\ref{eq:F8})].
}
\label{figH}
\end{figure}

\begin{figure}
\begin{center}
\end{center}
\caption{
The $s$ dependence of the inverse of the extended Fisher information 
$ \tilde{g}_q^{(2)}$ for various $q$ values 
with $N=2$ ($\mu=0.0$, $\sigma^2=1.0$).
}
\label{figJ}
\end{figure}

\begin{figure}
\begin{center}
\end{center}
\caption{
(Color online)
(a) The $N$ dependence of the Tsallis entropy per element,
$S_q^{(N)}/N$, for $q \leq 1.0$:
$(q,s)=(0.8, 0.0)$ (filled circles),  
$(0.8, 0.5)$ (filled squares),
$(0.9, 0.0)$ (open circles) and 
$(0.9, 0.5)$ (open squares).
(b) $S_q^{(N)}/N$ for $q \geq 1.0$:
$(q,s)=(1.01, 0.0)$ (filled circles),  
$(1.01, 0.5)$ (filled squares),
$(1.05, 0.0)$ (open circles) and 
$(1.05, 0.5)$ (open squares).
Dashed curves denote exact results 
given by Eq. (\ref{eq:B5})
[Figs. \ref{figE}(a) and \ref{figE}(b)] in appendix A.
Note the logarithmic and linear vertical scales in (a)
and (b), respectively.
}
\label{figF}
\end{figure}

\begin{figure}
\begin{center}
\end{center}
\caption{
The $N$ dependences of
inverses of the Fisher information matrices,
$g_q^{(1)}/g_q^{(N)}$
and $\tilde{g}_q^{(1)}/\tilde{g}_q^{(N)}$, 
for various $s$ values given by Eq. (\ref{eq:F7}): 
results for $s=-0.05$ and $s=-0.1$ are
valid for $N \leq 21$ and $N \leq 11$, respectively.
}
\label{figD}
\end{figure}

\begin{figure}
\begin{center}
\end{center}
\caption{
The probability distribution of $N=2$ systems, 
$p^{(2)}(x_1, x_2)$ [Eq. (\ref{eq:C5})],  
for (a) $s=0.0$ and (b) $s=0.5$ with $q=1.5$ 
as a function of $x_1$ for $x_2=0.0$, 0.5, and 1.0.  
}
\label{figC}
\end{figure}

\begin{figure}
\begin{center}
\end{center}
\caption{
(Color online)
The uncorrelated distribution  
for $N=1$, $p^{(1)}(x_1)$ [Eq. (\ref{eq:G7}): solid curves], and
marginal distributions of $p_m^{(2)}(x_1)$ for $N=2$ 
[Eq. (\ref{eq:G4}): dashed curves],
and $p_m^{(3)}(x_1)$ for $N=3$ [Eq. (\ref{eq:G6}): chain curves]
with $q=0.5$ and $q=1.5$ in (a) linear and (b) logarithmic vertical scales.  
}
\label{figB}
\end{figure}

\begin{figure}
\begin{center}
\end{center}
\caption{
The $N$ dependence of the Tsallis entropy per element,
$S_q^{(N)}/N$, in uncorrelated systems for (a) $q \leq 1.0$
and (b) $q \geq 1.0$:
note the logarithmic and linear vertical scales in (a)
and (b), respectively.
}
\label{figE}
\end{figure}

\end{document}